\documentclass[aps,twocolumn,superscriptaddress,nofootinbib,a4paper,longbibliography]{revtex4-2}

\usepackage{times}
\usepackage{epsfig}
\usepackage{amsfonts}
\usepackage{pifont}
\usepackage{amsmath}
\usepackage{amsthm}
\usepackage{amssymb}					
\usepackage{amsthm}
\usepackage{dsfont}
\usepackage{bm}
\usepackage{enumerate}
\usepackage{mathtools}
\usepackage{color}
\usepackage{multirow}
\usepackage[normalem]{ulem}
\newcommand{\stkout}[1]{\ifmmode\text{\sout{\ensuremath{#1}}}\else\sout{#1}\fi}
\usepackage{latexsym}
\usepackage{mathrsfs}
\usepackage{natbib}
\usepackage{verbatim}
\usepackage[T1]{fontenc}
\usepackage{float}
\usepackage{graphicx}
\usepackage{subcaption}
\usepackage{xcolor}
\usepackage{physics}
\usepackage{soul}
\usepackage{caption}
\usepackage{subcaption}
\usepackage{pifont}

\usepackage[font=small,labelfont=bf]{caption}

\usepackage{xspace}

\newcommand{\cmark}{\ding{51}}%
\newcommand{\xmark}{\ding{55}}%

\DeclarePairedDelimiter\floor{\lfloor}{\rfloor}

\usepackage{amssymb}
\usepackage{amsthm}
\usepackage{mathtools} 
\usepackage[customcolors]{hf-tikz} 
\usetikzlibrary{patterns}
\usetikzlibrary{matrix,decorations.pathreplacing}

\pgfkeys{tikz/mymatrixenv/.style={decoration={brace},every left delimiter/.style={xshift=8pt},every right delimiter/.style={xshift=-8pt}}}
\pgfkeys{tikz/mymatrix/.style={matrix of math nodes,nodes in empty cells,left delimiter={[},right delimiter={]},inner sep=1pt,outer sep=1.5pt,column sep=2pt,row sep=2pt,nodes={minimum width=20pt,minimum height=10pt,anchor=center,inner sep=0pt,outer sep=0pt}}}
\pgfkeys{tikz/mymatrixbrace/.style={decorate,thick}}

\usepackage[colorlinks=true,linkcolor=blue,citecolor=magenta,urlcolor=blue]{hyperref}

\begin{document}
	

\title{Binarisation-loophole-free observation of high-dimensional quantum nonlocality}

\author{Jia-le Miao}
\affiliation{Laboratory of Quantum Information, University of Science and Technology of China, Hefei 230026, China.}
\affiliation{CAS Center For Excellence in Quantum Information and Quantum Physics, University of Science and Technology of China, Hefei, 230026, China.}
\affiliation{Hefei National Laboratory, University of Science and Technology of China, Hefei 230088, China.}

\author{Elna Svegborn}
\affiliation{Physics Department and NanoLund, Lund University, Box 118, 22100 Lund, Sweden}

\author{Zhuo Chen}
\affiliation{Laboratory of Quantum Information, University of Science and Technology of China, Hefei 230026, China.}
\affiliation{CAS Center For Excellence in Quantum Information and Quantum Physics, University of Science and Technology of China, Hefei, 230026, China.}
\affiliation{Hefei National Laboratory, University of Science and Technology of China, Hefei 230088, China.}

\author{Yu Guo}
\affiliation{Laboratory of Quantum Information, University of Science and Technology of China, Hefei 230026, China.}
\affiliation{CAS Center For Excellence in Quantum Information and Quantum Physics, University of Science and Technology of China, Hefei, 230026, China.}
\affiliation{Hefei National Laboratory, University of Science and Technology of China, Hefei 230088, China.}

\author{Xiao-Min Hu}
\email{huxm@ustc.edu.cn}
\affiliation{Laboratory of Quantum Information, University of Science and Technology of China, Hefei 230026, China.}
\affiliation{CAS Center For Excellence in Quantum Information and Quantum Physics, University of Science and Technology of China, Hefei, 230026, China.}
\affiliation{Hefei National Laboratory, University of Science and Technology of China, Hefei 230088, China.}

\author{Yun-Feng Huang}
\affiliation{Laboratory of Quantum Information, University of Science and Technology of China, Hefei 230026, China.}
\affiliation{CAS Center For Excellence in Quantum Information and Quantum Physics, University of Science and Technology of China, Hefei, 230026, China.}
\affiliation{Hefei National Laboratory, University of Science and Technology of China, Hefei 230088, China.}

\author{Chuan-Feng Li}
\affiliation{Laboratory of Quantum Information, University of Science and Technology of China, Hefei 230026, China.}
\affiliation{CAS Center For Excellence in Quantum Information and Quantum Physics, University of Science and Technology of China, Hefei, 230026, China.}
\affiliation{Hefei National Laboratory, University of Science and Technology of China, Hefei 230088, China.}

\author{Guang-Can Guo}
\affiliation{Laboratory of Quantum Information, University of Science and Technology of China, Hefei 230026, China.}
\affiliation{CAS Center For Excellence in Quantum Information and Quantum Physics, University of Science and Technology of China, Hefei, 230026, China.}
\affiliation{Hefei National Laboratory, University of Science and Technology of China, Hefei 230088, China.}

\author{Armin Tavakoli}
\email{armin.tavakoli@teorfys.lu.se}
\affiliation{Physics Department and NanoLund, Lund University, Box 118, 22100 Lund, Sweden}

\author{Bi-Heng Liu}
\email{bhliu@ustc.edu.cn}
\affiliation{Laboratory of Quantum Information, University of Science and Technology of China, Hefei 230026, China.}
\affiliation{CAS Center For Excellence in Quantum Information and Quantum Physics, University of Science and Technology of China, Hefei, 230026, China.}
\affiliation{Hefei National Laboratory, University of Science and Technology of China, Hefei 230088, China.}
\affiliation{College of Physics, Guizhou University, Guiyang 550025, China}

\begin{abstract}
Bell inequality tests based on high-dimensional entanglement usually require measurements that can resolve multiple possible outcomes. However, the implementation of high-dimensional multi-outcome measurements is often  only emulated via a collection of ``click or no-click'' measurements. This reduction of multi-outcome measurements to binary-outcome measurements opens a loophole in high-dimensional tests Bell inequalities which can be exploited by local hidden variable models  [\href{https://doi.org/10.1103/PhysRevA.111.042433}{Tavakoli et al., Phys. Rev. A 111, 042433 (2025)}]. Here, we close this loophole by using four-dimensional photonic path-mode entanglement and multi-outcome detection. We test both  the well-known Collins-Gisin-Linden-Massar-Popescu inequality and a related Bell inequality  tailored for maximally entangled states in  high-dimension. We observe violations that are large enough to also rule out any quantum model based on entanglement of lower dimension, thereby demonstrating genuinely high-dimensional nonlocality free of the binarisation loophole.
\end{abstract}

\date{\today}

\maketitle

\textit{Introduction.---} Entanglement between quantum systems with more than two levels is called high-dimensional entanglement. It frequently  makes possible amplified quantum advantages, in for example cryptography \cite{Cerf2022}, entanglement distribution \cite{Horodecki1999} and steering \cite{Wiseman2007}, and therefore it has become an emerging resource in several areas of quantum information science. The realisation and control of high-dimensional entanglement has seen considerable progress in recent years \cite{Erhard2020}.  

The paradigmatic benchmark of entanglement is the violation of Bell inequalities \cite{Brunner_2014}. It is a phenomenon which in recent years has paved the way for areas such as quantum networks \cite{Tavakoli2022}, self-testing \cite{Supic2020} and device-independent quantum key distribution \cite{Zhang2022, Nadlinger2022}. More than 20 years ago, it was discovered that  $d$-dimensional entanglement can enhance the noise-tolerance of Bell inequality violations \cite{Collins_2002} and soon afterwards it was shown that non-maximally entangled states can further improve this feature \cite{Acin2002, Zohren2008}.  These so-called Collins-Gisin-Linden-Massar-Popescu (CGLMP)  inequalities, which require two measurements and $d$ possible outcomes per party, have remained the leading benchmark for Bell inequality tests beyond qubit dimension. Early violations of the CGLMP inequalities were based on three-level entanglement \cite{Vaziri2002, Thew2004} and several other demonstrations have followed \cite{Richart2012, Bernhard2013, Bessire2014, Ikuta2016, Lo2016, Cai2016, Zhang2024}. In particular, using photons entangled in orbital angular momentum, violations up to $d=12$ were reported in \cite{Dada_2011}. Violations have also been reported up to $d=8$ on integrated optical circuits \cite{Wang2018} using both the CGLMP test and other Bell tests tailored for maximally entangled states \cite{Son2006, Salavrakos2017}. In 2022, detection-loophole-free Bell inequality violation was achieved with high-dimensional entanglement \cite{Hu2022} and recently  high-dimensional Bell nonlocality was shown with  more than two parties \cite{Hu2025}.

\begin{figure}[t!]
        \includegraphics[width=0.45\textwidth]{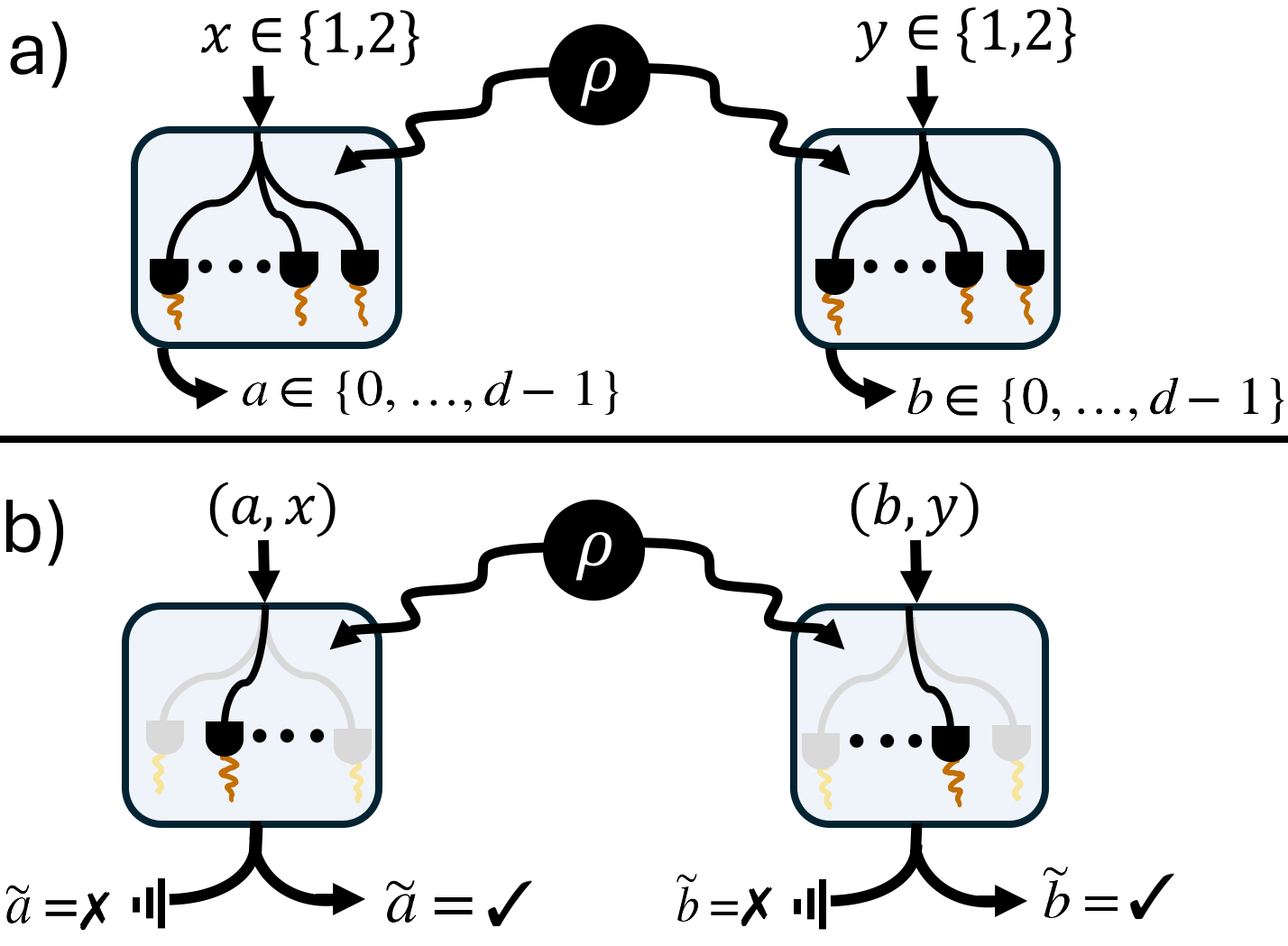}
    \caption{\textit{Multi-outcome vs binarised  CGLMP test.} Alice and Bob perform local measurements, $x$ and $y$, on a shared $d$-dimensional state $\rho$. (a) Multi-outcome measurements are used to resolve any one of the $d$ possible outcomes per party and round. (b) The multi-outcome measurements are emulated by a sequence of single-detector measurements, each oriented at the $a$'th ($b$'th) projection of the $x$'th ($y$'th) measurement. By post-processing the relative frequency of successful projections, one estimates the statistics of the multi-outcome measurement used in the Bell test.} \label{fig:multi_bin}
\end{figure}

However, these Bell inequality tests for $d$-dimensional entanglement do not implement the multi-outcome measurements stipulated in the theoretical description of the test. The CGLMP inequality presumes that in every round of the experiment, any one of the $d$ possible outcomes can be registered for Alice and Bob respectively; see Fig~\ref{fig:multi_bin}a.
In contrast, such multi-outcome measurements are typically implemented by an emulation procedure based on a collection of ``click or no-click'' measurements. In these, a click (\cmark) corresponds to a projection onto the outcome $\ket{i}$ and a no-click (\xmark) corresponds to the failure of said projection;  see Fig~\ref{fig:multi_bin}b. This binary-outcome measurement is repeated for each $i\in\{0,\ldots,d-1\}$ and by post-processing the data obtained from the $d$ distinct measurements, one estimates the statistics of the intended multi-outcome measurement, $\{\ket{i}\}_{i=0}^{d-1}$. It has been brought to attention that such a binarised implementation of a multi-outcome measurement opens a loophole in Bell inequality tests, as well as several other types of quantum correlation tests \cite{Tavakoli_2025}. The cause of the loophole is that Bell tests are not supposed to adopt any physical model of the measurement devices. In contrast, a binarised implementation of a multi-outcome measurement not only implies assuming the Hilbert space dimension, but also the orthogonality of the projections implemented by different settings. Making such an unwarranted assumption is  at odds with the nature of a Bell test and therefore leads to the opening of the  binarisation loophole. This is relevant independently of whether  one has closed the locality and detection loopholes, which are well-known from Bell tests that rely on standard qubit entanglement. 


Here we employ four-dimensional path-mode entanglement and perform  multi-outcome measurements to close the binarisation loophole in Bell experiments. Our multi-outcome measurement device enables flexible switching of the projection basis, yielding sizable violations of both the CGLMP inequality using non-maximally entangled states and related Bell inequalities tailored for maximally entangled states. The violations  significantly exceed the the amounts of nonlocality possible with any lower-dimensional entanglement, thereby demonstrating a genuinely high-dimensional phenomenon.

\textit{The binarisation loophole.---}
The binarisation loophole arises in Bell tests when multi-outcome measurements are implemented via a set of binary-outcome measurements \cite{Tavakoli_2025}. Consider a Bell experiment in which Alice and Bob input $(x,y)$, and their outputs $(a,b)$ have $d$ possible values each. The correlations observed in this scenario, when measuring the shared state $\rho$, are described by the probability distribution $p_{\text{multi}}(a,b | x,y)=\Tr\left(A_{a|x}\otimes B_{b|y}\rho\right)$. In a binarised implementation, Alice's measurement projectors, $A_{0|x},..., A_{d-1|x}$, are replaced with a sequence of $d$ projective measurements. Each of these, which we denote by $\{\tilde{A}_{\tilde{a} \mid a,x}\}_{\tilde{a}}$, is  a binary-outcome measurement. The possible outcomes, labeled $\tilde{a}$, are either a successful ($\tilde{a}$ = \cmark) or failed ($\tilde{a}$ = \xmark) projection onto the $a$'th multi-outcome measurement projector $A_{a\mid x}$; see Figure \ref{fig:multi_bin}. That is, the successful outcome is intended to represent $\tilde{A}_{\text{\cmark} \mid a,x}=A_{a \mid x}$, and the unsuccessful one its complement. Similarly, Bob's binarised measurement operators are denoted by $\tilde{B}_{\tilde{b} \mid b,y}$, with $\tilde{b} \in \{\text{\cmark, \xmark}\}$. The binarised probability distribution then takes the form
\begin{equation}\label{binarised}
p_{\text{bin}}(\tilde{a},\tilde{b} \mid (a,x), (b,y)) = \Tr \left( \tilde{A}_{\tilde{a} \mid a,x} \otimes \tilde{B}_{\tilde{b} \mid b,y} \rho \right). 
\end{equation}

By post-processing the relative frequency of successful projections in the binarised implementation, one reconstructs the multi-outcome measurement statistics, $p_{\text{multi}}(a,b| x,y) = p_{\text{bin}}(\text{\cmark},\text{\cmark} | (a,x), (b,y))$. However, this substitution assumes that the set of $d$ independent projections $\{ \tilde{A}_{\text{\cmark}  \mid a,x} \}_{a}$ forms a valid measurement, i.e.~that it satisfies the normalisation condition $\sum_a \tilde{A}_{\text{\cmark}  \mid (a,x)} = \openone$. This assumption is not justified in a  Bell test because each physical setting, which in the binarised implementation is the pair $(a,x)$ for Alice (and similarly for Bob), must be treated as uncharacterised. Consequently, by introducing this unwarranted constraint in a Bell experiment, one opens a loophole that can be exploited by local hidden variable models to account for the apparent nonlocality.

\begin{figure}[t!]
    \centering
    \includegraphics[width=\linewidth]{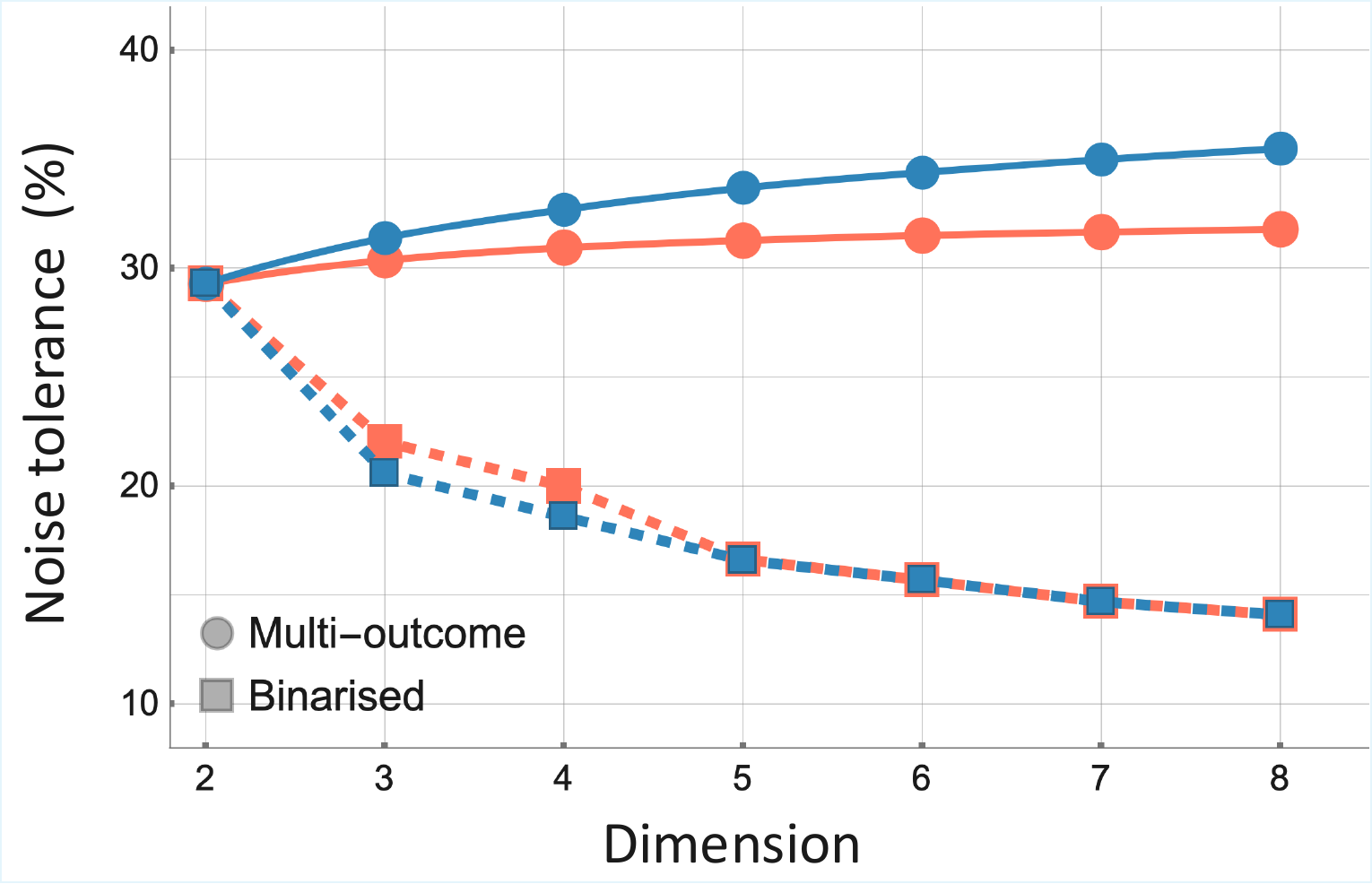}
    \caption{White noise tolerance of Bell nonlocality in the tests associated to $\mathcal{I}_4$ (blue) and $\mathcal{S}_4$ (red). Multi-outcome measurements (circular marker) have stronger robustness in larger dimension whereas binarised measurements (rectangular marker) have the opposite trend.}
    \label{fig:highDim}
\end{figure}

\textit{High-dimensional Bell tests.---}
In the Bell scenario considered in  the CGLMP test, two distant parties, Alice and Bob, independently select binary inputs, $x,y \in \{1,2\}$, and perform associated measurements, whose outcomes are denoted $a,b \in \{0,...,d-1\}$.  The resulting input/output statistics is characterised by the joint probability distribution $p_\text{multi}(a,b|x,y)$. If the probability distribution is compatible with a local hidden variable model, it must satisfy the CGLMP inequality \cite{Collins_2004, Zohren2008}
\begin{equation}
\begin{aligned}
\mathcal{I}_d &= P(A_1 \leq B_1) + P(B_1 \leq A_2) \\
&+ P(B_2 \leq A_1) - P(B_2 \leq A_2) - 2 \leq 0,
\end{aligned}
\end{equation}
 where $P(A_x \leq B_y) \equiv \sum_{a \leq b} p_\text{multi}(a,b | x,y)$. For $d=2$ it reduces to the well-known CHSH inequality, but for $d>2$ it achieves its maximal violation with a non-maximally entangled $d$-dimensional  state. Given the general interest in the maximally entangled state, it was later shown that one can  modify the CGLMP inequality so that the maximal quantum violation is achieved with the maximally entangled $d$-dimensional state \cite{Son2006, Salavrakos_2017}. We write this other Bell inequality as $\mathcal{S}_d\leq 0$, and refer to Appendix \ref{App:DW} for its details. 

\begin{figure*}[ht!]
	\includegraphics[width=0.75\textwidth]{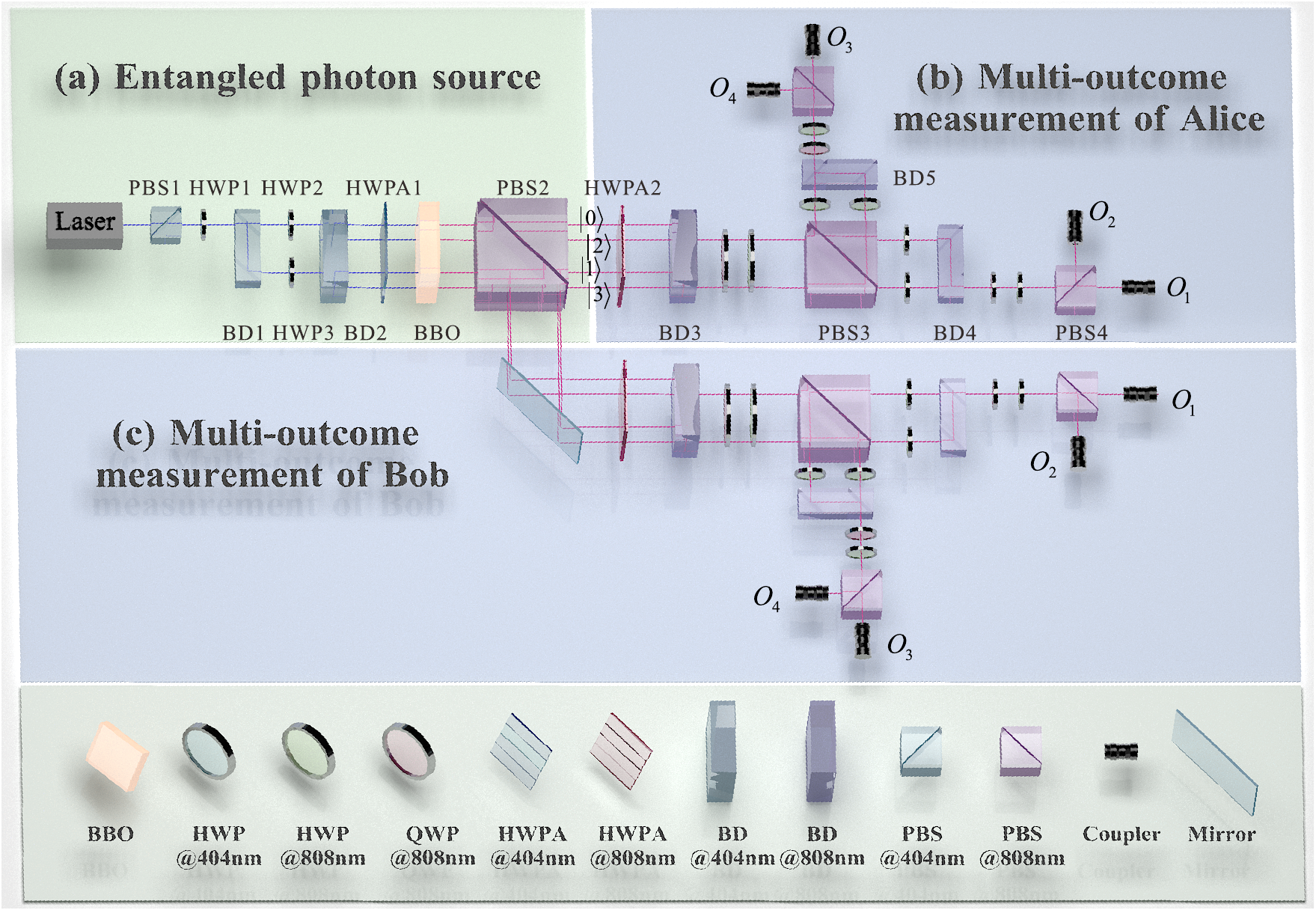}
	\caption{Experimental setup. (a) Photon Source. A continuous-wave laser operates at 404~nm is split into parallel paths with 4~mm spacing in the horizontal direction after passing through BD1 and BD2, and 4~mm spacing vertically. When the laser array passes through the BBO crystal, SPDC process  generates photon pairs (signal and idler) entangled in path DOF. Then the photon pairs are split by PBS2, sent to Alice and Bob. (b),(c) Multi-outcome measurements of Alice and Bob. Depending on their inputs $x\in\{1, 2\}$ (Alice) and $y\in\{1, 2\}$ (Bob), each party performs a local four-outcome measurement. First, HWPA2—an array of 808~nm half wave plates set at $0^\circ$ or $45^\circ$ in sequence—adjusts the polarisation in each path. The initial 2×2 path arrangement is then combined by BD3 and BD4 into two paths, converting path encoding into polarisation-path hybrid encoding. The measurement finally projects onto four different outcomes, collected separately by couplers $O_1$ to $O_4$. Coincidence events between all of Alice’s and Bob’s couplers are registered by a time to digital converter (UQDevice) with a coincidence window of 3~ns. HWP: Half wave plate. QWP: Quarter wave plate. HWPA: Half wave plate array. BD: Beam displacer. PBS: polarising beam splitter. BBO: $\beta$-barium-borate crystal.}
	\label{fig:setup}
	\vspace{0.0cm}
\end{figure*}

Many times, standard qubit entanglement is sufficient to violate a Bell inequality tailored for high-dimensional entanglement. Therefore, it is important to observe  large enough values of both  $\mathcal{I}_d$ and $\mathcal{S}_d$ to show not only nonlocality but genuinely high-dimensional nonlocality. For the latter, one must prove that the distribution $p_\text{multi}(a,b|x,y)$ admits no quantum model based on lower-dimensional entanglement, i.e.~there exists no possible quantum measurements $\{A_{a|x}\}$ and $\{B_{b|y}\}$ and no possible entangled state $\rho$ of dimension $D<d$ such that  $p_\text{multi}(a,b|x,y)  = \tr(A_{a\mid x} \otimes B_{b\mid y} \rho)$ up to convexification. Since our experiment focuses on $d=4$, we compute bounds on the Bell parameters $\mathcal{I}_4$ and $\mathcal{S}_4$ for  $D=1,2,3,4$ (note that $D=1$ is the same as local hidden variable models),
\begin{equation}\label{eq:upper_bounds}
\begin{aligned}
&\mathcal{I}_4 \stackrel{\text{LHV}}{\leq} 0  \stackrel{\text{D} = 2}{\leq} 0.207   \stackrel{\text{D} = 3}{\leq} 0.305   \stackrel{\text{D} = 4}{\leq} 0.365\\[1ex]
&\mathcal{S}_4 \stackrel{\text{LHV}}{\leq} 0  \stackrel{\text{D} = 2}{\leq} 0.152   \stackrel{\text{D} = 3}{\leq} 0.212   \stackrel{\text{D} = 4}{\leq} 0.302.
\end{aligned}
\end{equation}
The bounds are obtained from semidefinte relaxation methods \cite{TavakoliSDP2024} following the technique of  Ref~\cite{Navascues_2015}; see Appendix \ref{App:Corr_ext}. As seen above, demonstrating genuine four-dimensional nonlocality is more demanding than basic nonlocality. One can estimate this difference by considering  the mixture of the optimal entangled state with white noise. Violating the LHV-bound for $\mathcal{I}_4$ ($\mathcal{S}_4$) requires a visibility of $67.3\%$ ($69.1\%$) whereas violating the bound for $D=3$  requires a visibility of at least $94.4\%$ ($94.0\%$).

Consider now that the Bell tests associated to $\mathcal{I}_d$ and $\mathcal{S}_d$ are instead implemented using binarised measurements for Alice and Bob. The experimental distribution will not be the desired $p_\text{multi}$ but instead $p_\text{bin}$ of the form in Eq~\eqref{binarised}. As shown in \cite{Tavakoli_2025} for $\mathcal{I}_d$ and in Appendix~\ref{App:bin} for $\mathcal{S}_d$, a binarised experiment still eludes a local hidden variable model. However, the magnitude of the violations, measured in terms of their white noise tolerance, decreases as we increase the dimension. This is a trend opposite to that encountered for multi-outcome Bell tests (see Figure~\ref{fig:highDim}) and it partly defeats the motivation for employing high-dimensional systems. Furthermore, in Appendix~\ref{App:bin} we determine the optimal Bell inequality for detecting the nonlocality of $p_{\text{bin}}$ associated with the binarised implementation of $\mathcal{I}_4$ and  $\mathcal{S}_4$. We  show that not only do the binarised Bell tests lose the noise advantage of higher dimension, but they also lose the capability to detect genuine four-dimensional nonlocality.






\textit{Experiment.---} We present an experimental demonstration of genuine high-dimensional quantum nonlocality free from the binarisation loophole. Our setup is illustrated in Fig~\ref{fig:setup}. It features a source of four-dimensional photonic entanglement encoded in path degree of freedom and four-outcome measurements for both Alice and Bob. We conduct separate tests of the two inequalities in \eqref{eq:upper_bounds}. For these, we prepare the optimal  state, which for $\mathcal{S}_4$ is maximally entangled and for $\mathcal{I}_4$ is partially entangled. 

In the stage of state preparation,  a CW laser at $404$~nm is first polarised by PBS1 and then separated into a $2\times 2$ array of parallel beams with a spacing of $4$mm both horizontally and vertically. The BD1 (BD2) refracts horizontally (vertically) polarised beams in vertical (horizontal) direction by 4mm.  After the separation, a 4mm-thick $\beta$-barium borate crystal is illuminated by the array of beams, and the two photon path-entangled state is then generated by SPDC process, which reads
$\ket{\psi_{AB}}=\sum\limits_{k=0}^{3}\lambda_i\ket{k_H k_V}$ with two photons in different polarisation. After the pump laser filtered by a long-pass filter and a 3nm band-pass filter centered at 808 nm, the two entangled photons are separated by PBS2 and sent to Alice and Bob for their multi-outcome measurement. The coefficients $\lambda_k$ are controlled by HWP1-3, so we can prepare both the maximal and non-maximal entangled state needed, i.e. $\lambda_0=\lambda_1=\lambda_2=\lambda_3=1/2$ for $\mathcal{S}_4$ inequality and $\lambda_0=\lambda_3=0.5686$, $\lambda_2=\lambda_4=0.4204$ for the $\mathcal{I}_4$ inequality. 

The construction of multi-outcome measurement setups has consistently posed significant challenges, typically requiring unitary transformations to achieve multi-outcome detection. To simplify the optical configuration, our scheme realizes high-dimensional multi-outcome measurement through sequential projections onto two-dimensional subspaces. As shown in Fig.~\ref{fig:setup}, during the measurement process, the path entanglement is converted stepwise into path-polarization entanglement via HWPAs and BDs, and then projected into different subspaces using a PBS. First, the polarisation of photons in different paths is adjusted by HWPA2. The upper and lower pairs of paths are then recombined by BD3, which refracts horizontally polarised photons downward by $4~mm$. As a result, the subspaces originally encoded in separate paths are now in the same spatial path but distinguished by polarisation.

Next, the relative phases between the subspaces $\{ \ket{0}, \ket{2} \}$ and $\{ \ket{1}, \ket{3} \}$ are measured using a combination of QWP, HWP, and PBS. With the QWP and HWP set at specific angles, photons transmitted or reflected by PBS3 are projected onto two distinct subspaces, differing only in the relative phase between the two measured subspace pairs. Specifically, photons in $\{ \ket{0} + e^{i\varphi_{02}} \ket{2}, \ket{1} + e^{i\varphi_{13}} \ket{3} \}$ are transmitted, while those in $\{ \ket{0} - e^{i\varphi_{02}} \ket{2}, \ket{1} - e^{i\varphi_{13}} \ket{3} \}$ are reflected. These two subspaces remain encoded in two separate paths with a horizontal spacing of 4~mm. This process is repeated once more: BD4 and BD5 combine the two polarisation-distinguished paths in the horizontal direction, refracting vertically polarised photons sideways by 4~mm. The relative phases between $\{ \ket{0}, \ket{1} \}$ and $\{ \ket{2}, \ket{3} \}$ are then measured, resulting in four possible output paths for the photons, each corresponding to one of the four basis states of an operator. Since photons corresponding to different measurement outcomes are spatially separated, Alice and Bob can simultaneously collect photon counts for all four outcomes using the couplers $O_1$--$O_4$. 

Photons are collected by couplers no matter what the measurement result is; none of the photons are post-selected and the data is normalised directly between the counts of the four couplers. The optimal measurements for the Bell tests are given in Appendix~\ref{App:DW} and they are realised by adjusting the angles
of QWPs and HWPs for Alice and Bob respectively. In general, our setup is capable of switching between any Fourier basis and the computational basis. The angles of the wave plates and the matching between $O_i$ and the measurement bases are found in the Appendix~\ref{sec:exp_measurement}.


\begin{figure}[t!]
    \centering
    \includegraphics[width=\linewidth]{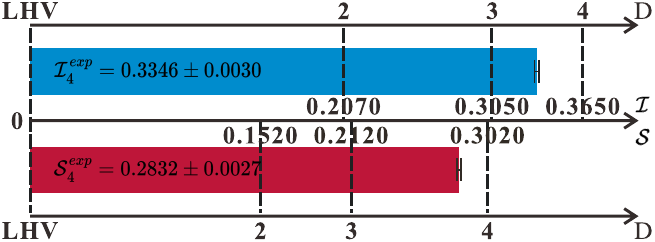}
    \caption{Experiment results for $\mathcal{I}_4$ (blue) and $\mathcal{S}_4$ (red), and upper bounds on quantum nonlocality for different dimensions. In both cases, we violate the Bell inequality and the limitations of three-dimensional  entanglement. 
    The statistical error for each violation is displayed at the end of each bar.}
    \label{fig:resultsbar}
\end{figure}

\textit{Results.---} The data is collected in 100~s per setting of Alice and Bob, with coincidence photon pair of approximately 400 Hz in total. The coincidence counts are normalised for each input pair $(x, y)$ so that we can estimate a $4\times 4$ joint probability distribution $P(x, y)$.  These measured values are given in Appendix~\ref{sec:exp_data}. The observed Bell parameters are 
\begin{equation}
	\begin{aligned}
		& \mathcal{I}_4^{\text{exp}}=0.3346\pm0.0030,\\ &\mathcal{S}_4^{\text{exp}}=0.2832\pm0.0027,
	\end{aligned}
	\label{eq:expresult}
\end{equation}
which are also illustrated in Fig. \ref{fig:resultsbar}. Both exceed the bounds for LHVs and qutrit entanglement in Eq~\eqref{eq:upper_bounds}. The latter violations correspond to $9$ and $26$ standard deviations respectively. In Appendix~\ref{App:Counts} we show that it corresponds to a negligible statistical uncertainty.

The result \eqref{eq:expresult} indicates that we have reached a visibility of  $97.30\%$ for the non-maximal entangled state and $98.80\%$ for the maximal entangled state, when considering the white noise model. We have also estimated the measurement fidelity from our data by assuming the state is ideal (quantum state fidelity equals to 1) but measurement projectors are mixed with white noise. Based on this model and our experimental data, the estimated average measurement fidelity is approximately $98.8\%$ for $\mathcal{I}_4$ and $99.3\%$ for $\mathcal{S}_4$.


\textit{Discussion.---}
We have demonstrated nonlocality from high-dimensional entanglement via the implementation of multi-outcome measurements on the path-modes of single photons. In this way, we have closed the binarisation loophole for high-dimensional Bell tests.   Our results cover both the seminal CGLMP test and more recent Bell inequalities tailored for high-dimensional maximally entangled states. In both cases, we observe Bell inequality violations sufficiently large that they cannot be modelled with lower-dimensional entanglement, thereby showcasing a genuinely high-dimensional effect.

Our results take further the recent efforts to close the binarisation loophole in high-dimensional quantum information. It was recently closed in a dense-coding-inspired quantum communication experiment  \cite{Zhang2025}. This test has the advantage that the  quality benchmarks required for observing genuinely high-dimensional effects are lower, but the drawback of being conceptually weaker form of non-classicality than Bell nonlocality, without a pathway to device-independence. 
In addition, Ref~\cite{Dekkers2025} recently reported binarisation-loophole-free violations of the CGLMP inequality via  photonic spectral measurements. In this experiment, Alice and Bob cannot actively choose their measurement settings, and the reported Bell violations are too small to go beyond what is in principle achievable with standard two-qubit entanglement. Both these limitations are overcome in our present demonstration.

Our work underlines the  role of multi-outcome measurements as a key enabling technology for high-dimensional tests of quantum theory and its associated quantum information applications. A central challenge is that of making multi-outcome measurements scalable. A promising path for that is integrated optics \cite{Wang2018}. On-chip integration of single-photon detectors can also facilitate the next open challenge of closing the detection loophole \cite{li2025surpassing} in a binarisation-loophole-free high-dimensional Bell test.

\begin{acknowledgments}
This work was supported by the Quantum Science and Technology-National Science and Technology Major Project (No. 2024ZD0301400, No. 2021ZD0301200), the NSFC (No. U25D8007, No. 62322513, No. 12374338,  No. 12204458, and No. 12350006), the Fundamental Research Funds for the Central Universities, Anhui Provincial Natural Science Foundation (No. 2408085JX002), Anhui Province Science and Technology Innovation Project (No. 202423r06050004). E.S and A.T are supported by the Knut and Alice Wallenberg Foundation through the Wallenberg Center for Quantum Technology (WACQT), the Swedish Research Council under Contract No.~2023-03498 and the Swedish Foundation for Strategic Research.
\end{acknowledgments}

	\twocolumngrid
		\bibliography{references}
	
\clearpage

\appendix

\section{High-dimensional Bell tests}\label{App:DW}
Here we introduce in detail the CGLMP inequality and its related variant tailored for maximally entangled states. We discuss how both of these witnesses can be used to certify nonlocality that is compatible only with high-dimensional entanglement.

\subsection{CGLMP inequality}
Consider the CGLMP scenario in which Alice and Bob select binary inputs $x,y \in \{1,2\}$,  and perform associated measurements on a bipartite state $\rho \in \mathbb{C}^D \otimes \mathbb{C}^D$. The generated outputs take values $a,b \in \{0,...,d-1\}$. For any integer $d\geq 2$ a facet of the local polytope is given by the CGLMP inequality \cite{Collins_2004,Acin_2006}
\begin{equation}\label{eq:CGLMP}
\begin{aligned}
\mathcal{I}_d &= P(A_1 \leq B_1) + P(B_1 \leq A_2) \\
&+ P(B_2 \leq A_1) - P(B_2 \leq A_2) - 2 \leq 0,
\end{aligned}
\end{equation}
where we have defined $P(A_x \leq B_y) \equiv \sum_{a \leq b} p(a,b | x,y)$. To maximally violate the inequality in quantum theory, Alice and Bob perform the measurements 
\begin{equation} \label{eq:opt_meas}
    \begin{aligned}
        &|A_{a\mid x} \rangle = \dfrac{1}{\sqrt{d}}\sum_{k=0}^{d-1}e^{\frac{i 2\pi}{d} k ( a+\alpha_x)}|k\rangle_A \\
        &|B_{b\mid y} \rangle = \dfrac{1}{\sqrt{d}}\sum_{k=0}^{d-1}e^{\frac{i 2\pi}{d}k (- b+\beta_y)}|k\rangle_B
    \end{aligned}
\end{equation}
with the phases $\alpha_1 = 0$, $\alpha_2 = 1/2$, $\beta_1 = -1/4$ and $\beta_2 = 1/4$. For $d=2$  the CGLMP inequality reduces to the CHSH inequality, and is maximally violated by a maximally entangled state. For higher dimensions, namely $d>2$, the optimal $d$-dimensional entangled state is partially entangled \cite{Zohren2008}. 

Consider the case of $d=4$, which is the focus of our experiment. The optimal four-dimensional entangled state reads
\begin{equation}\label{eq:opt_state}
\ket{\psi_4} = \lambda_0\ket{00} + \lambda_1 \ket{11} + \lambda_2 \ket{22} + \lambda_3 \ket{33}.
\end{equation}
where $\lambda_0 = \lambda_3 =0.5686$ and $\lambda_1 = \lambda_2 = 0.4204$ \cite{Cai_2016}. The optimal state and measurements yield that CGLMP value
\begin{equation}\label{eq:4_4}
\mathcal{I}_4 = 0.365.
\end{equation}
In contrast, if the shared state $\rho$ is restricted to a three-dimensional entangled state, $D = 3$, whereas the number of possible measurement outcomes remains $d = 4$, the quantum violation of the CGLMP inequality can at most reach
\begin{equation}\label{eq:4_3}
\mathcal{I}_4 \leq 0.305.
\end{equation}
This bound was conjectured to be optimal in Ref~\cite{Cai_2016} and in Appendix~\ref{App:Corr_ext} we prove the conjecture.

The value in Eq~\eqref{eq:4_4} can be saturated under ideal circumstances. However, in practice no experiment is noiseless. Therefore, it is interesting to compute the critical noise-limit at which we no longer can distinguish whether a violation of the four-outcome CGLMP inequality is due to a $4$-dimensional or a $3$-dimensional entangled state. Consider that with probability $v$ we generate the optimal four-dimensional entangled state $\Psi_4 = \ketbra{\psi_4}{\psi_4}$, and with probability $1-v$ we generate the maximally mixed state
\begin{equation}\label{eq:rho_v}
\rho = v\Psi_4 + (1-v) \frac{\mathds{1}}{16}.
\end{equation}
Using that the CGLMP inequality is linear, we find that the critical noise tolerance, required to distinguish four- and three-dimensional entanglement is 
\begin{equation}
    v_{\text{crit}} = 0.946.
\end{equation}

\subsection{Bell inequality for maximally entangled state}
We now consider a Bell inequality related to the CGLMP inequality, which is tailored for high-dimensional maximally entangled states \cite{Son2006, Salavrakos_2017}. Like CGLMP, it has two inputs per party and $d$-outcomes per party. The relevant correlation parameter is  
\begin{equation}
\sum_{k=0}^{\floor*{d/2}-1}  (\alpha_k \mathbb{P}_k - \beta_k \mathbb{Q}_k).
\end{equation}
where the coefficients are given by
\begin{equation}
\begin{aligned}
&\alpha_k = \frac{1}{2d} \tan(\frac{\pi}{4}) \bigg[g(k)-g\big(\floor{\frac{d}{2}}\big)\bigg], \\
&\beta_k = \frac{1}{2d} \tan(\frac{\pi}{4}) \bigg[g(k+\frac{1}{2})-g\big(\floor{\frac{d}{2}}\big)\bigg], 
\end{aligned}
\end{equation}
with $g(x) = \cot(\pi(x+1/4)/d)$. Moreover, the correlators of interest are defined as 
\begin{equation}
\begin{aligned}
&\mathbb{P}_k \equiv \sum_{i = 1}^2 \big[P(A_i = B_i + k) + P(B_i = A_{i+1}+k)\big], \\
&\mathbb{Q}_k \equiv \sum_{i = 1}^2 \big[P(A_i = B_i - k-1) + P(B_i = A_{i+1}-k-1)\big], 
\end{aligned}
\end{equation}
with $A_{3} \equiv A_1 + 1$. Here, we have used the notation
\begin{equation}
P(A_x = B_y + k) \equiv \sum_{j=0}^{d-1} p(j + k, j |x,y ).
\end{equation} 
where addition is taken modulo $d$. The inequality is maximally violated when Alice and Bob perform optimal CGLMP measurements, defined in Eq~\eqref{eq:opt_meas}, with phases $\alpha_x = (x-1/2)/2$ and $\beta_y = y/2$, on the two-qudit maximally entangled state
\begin{equation}
\ket{\phi^+} = \frac{1}{\sqrt{d}} \sum_{i=0}^{d-1}\ket{ii}.
\end{equation}

Consider now $d = 4$. The Bell inequality can be written 
\begin{equation}
\mathcal{S}_{4} = \sum_{k=0}^{1}  (\alpha_k \mathbb{P}_k - \beta_k \mathbb{Q}_k) - 1.798 \leq 0,
\end{equation}
where the constant is infered from local hidden variable models.  By sharing a four-dimensional maximally entangled state, the largest violation becomes
\begin{equation}
\mathcal{S}_{4} =0.3019.
\end{equation}
In contrast, the largest violation possible with three-dimensional entanglement is in Appendix~\ref{App:Corr_ext} proven to be bounded as
\begin{equation}\label{satwap}
\mathcal{S}_{4} \leq 0.2117.
\end{equation}
Hence, we see that $\mathcal{S}_{4}$ can reveal genuine four-dimensional nonlocality. By computing the white noise tolerance required to distinguish the entanglement dimension $D = 4$ from $D = 3$, we find that
\begin{equation}
v_{\text{crit}} = 0.940.
\end{equation}

\section{Rank constrained optimisation in a Bell scenario}\label{App:Corr_ext}
In this section we show how to compute the bounds in Eq~\eqref{eq:4_3} and Eq~\eqref{satwap} for three-dimensional entanglement in the two relevant Bell tests. To this end, we use the the heuristic semidefinite programming method of Ref~\cite{Navascues_2015}.

Let $\rho$ be the shared $D$-dimensional state and let  $\{A_{a \mid x}\}_a$ and $\{B_{b \mid y}\}_b$ be the local projective measurements of Alice and Bob. 
A $d$-outcome projective measurement over a Hilbert space of dimension $D<d$ must have at least $d-D$ outcome associated with zero-projectors. We associate each of Alice's measurements with a rank-vector 
\begin{equation}
\vec{r}_{A}^{\ (x)} = (r_0^{(x)},r_1^{(x)},...,r_{d-1}^{(x)}),
\end{equation}
where we define $r_a^{(x)} \equiv \text{rank}(A_{a\mid x})$. Here, every combination of $\vec{r}_{A}^{ \ (x)} \in \{0,1\}^4$, such that $\sum_a r_a^{ \ (x)} = D$, is a valid tuple. Similarly, we introduce the rank-vector $\vec{r}_B^{\ (y)}$ for Bob. For each valid pair of rank-combination $(\vec{r}_{A}^{\ (x)},\vec{r}_{B}^{\ (y)})$, we bound the maximal violation of the CGLMP inequality using the heuristic sampling method proposed in \cite{Navascues_2015}. For completeness, we include a summary of the method below.

\subsection{Heuristic sampling method}
Consider that Alice and Bob perform $d$-outcome measurements $\{A_{a\mid x}\}_a$ and $\{B_{b\mid y}\}_b$. For each measurement setting $x,y$, we assume that $D$ of the measurement outcomes $a,b$ are associated with a rank-one projector whereas the remaining are associated with zero projectors. We encode this information into the pair of rank-vectors $(\vec{r}_{A}^{\ (x)},\vec{r}_{B}^{\ (y)})$. In accordance with the rank-vectors we then define an operator list $L= \{ \openone, \{A_{a\mid x}\}_a, \{B_{b\mid y}\}_b \}$ \cite{Navascues_2015}. Here, Alice's and Bob's rank-one projectors are generated by sampling randomly over $\mathbb{C}^D$ while the remaining are set to zero projectors. Over $L$ we define a monomial list $\mathcal{S}$, of least at length one, meaning that $\mathcal{S}$ must at least include all elements in $L$ \cite{Navascues_2015}. From the monomial list we then define the $|\mathcal{S}| \times |\mathcal{S}|$ moment matrix
\begin{equation}
\Gamma_{uv} = \tr(u^\dagger v \rho)
\end{equation}
for $u,v \in \mathcal{S}$ and some $D$-dimensional quantum state $\rho$. By construction we have that $\Gamma \succeq 0$. We then repeat this procedure independently and denote each generated moment matrix by $\Gamma_k$. The iteration is terminated when the next moment matrix is linearly dependent of the previously generated samples $\{\Gamma_k\}_k$. The final moment matrix is then defined as an affine combination over the samples, i.e.,
\begin{equation}
\begin{aligned}
\Gamma = \sum_k s_k \Gamma_k
\end{aligned}
\end{equation}
for some coefficients $\{s_k\}_k$ satisfying $\sum_k s_k = 1$ \cite{Navascues_2015}. Over this domain we then optimise the quantum violation of the relevant Bell inequality, namely,
\begin{equation}\label{eq:mm_CGLMP}
\mathcal{I}_{d} = \sum_{a b x y} c_{abxy} \Gamma_{A_{a\mid x}, B_{b \mid y}}.
 \end{equation} 
By solving this SDP we obtain an upper bound of the Bell inequality. We then repeat this procedure for each valid rank-combination $(\vec{r}_{A}^{\ (x)},\vec{r}_{B}^{\ (y)})$. After solving the problem separately for each rank-combination, the define our rank-constrained upper bound $\mathcal{I}_{d}^*$ to the Bell inequality as follows
\begin{equation}
\mathcal{I}_{d}^* = \max_{\vec{r}_{A}^{\ (x)},\vec{r}_{B}^{\ (y)}} \quad \mathcal{I}_{d}.
\end{equation}
By small modifications, this method can also be used to compute the optimal violation achievable with qubit-entanglement given in the main text.

\subsection{Optimal rank-vectors}
Lastly, we include the optimal rank-vector that yields maximal valuation $\mathcal{I}_4 \leq 0.305$ of the CGLMP inequality with qutrit measurements acting on a qutrit state. 
\begin{equation}
\begin{aligned}
&\vec{r}_A^{\ (1)} = (1,0,1,1), \quad \vec{r}_A^{\ (2)} = (0,1,1,1),\\
&\vec{r}_B^{\ (1)} = (0,1,1,1), \quad \vec{r}_B^{\ (2)} = (1,0,1,1).
\end{aligned}
\end{equation}
We also include the rank-vectors that provide the upper bound $\mathcal{S}_{4} \leq 0.2117$ for the Bell inequality tailored for maximally entangled states
\begin{equation}
\begin{aligned}
&\vec{r}_A^{\ (1)} = (1,1,0,1), \quad \vec{r}_A^{\ (2)} = (1,1,0,1),\\ 
&\vec{r}_B^{\ (1)} = (1,1,0,1), \quad \vec{r}_B^{\ (2)} = (1,1,1,0).
\end{aligned}
\end{equation}

\section{Binarisation procedure}\label{App:bin}
We analyse the quantum properties of the correlations obtained in a binarised Bell experiment, following Ref~\cite{Tavakoli_2025}. Consider that in the multi-outcome scenario, Alice and Bob, independently select input $(x,y) \in \{1,2\}$, perform associated measurements on a shared state $\rho$ with outputs $(a,b) \in \{0,...,d-1\}$. Now, assume that both Alice and Bob binarise their measurements. Their respective input to the binarised measurements are then described by the tuples $\tilde{x} = (a,x)$ and $\tilde{y} = (b,y)$, respectively. The associated probability distribution $p_{\text{bin}}$ is given by Eq~\eqref{binarised}. This set of correlations $p_{\text{bin}}$ can then be expressed in terms of the multi-outcome measurements as follows
\begin{equation}\label{eq:bin_rel}
\begin{aligned}
        &p_{\text{bin}}(0,0\lvert \tilde{x}, \tilde{y})  = \tr(A_{a \lvert x} \otimes B_{b \mid y} \rho)\\
        &p_{\text{bin}}(0,\perp\lvert \tilde{x}, \tilde{y})  = \tr(A_{a \mid x} \rho^A) -  \tr(A_{a\lvert x} \otimes B_{b \lvert y} \rho)\\
        &p_{\text{bin}}(\perp ,0\lvert\tilde{x}, \tilde{y})  = \tr(B_{b \mid y} \rho^B)-  \tr(A_{a\lvert x} \otimes B_{b \lvert y} \rho)\\
        &p_{\text{bin}}(\perp,\perp\lvert\tilde{x}, \tilde{y})  = 1 -  \tr(A_{a \mid x} \rho^A) -  \tr(B_{b \mid y} \rho^B) \\
        & \qquad \qquad \qquad \quad \ + \tr(A_{a \mid x} \otimes B_{b \mid y} \rho)
\end{aligned}
\end{equation}
Here, $\rho^{A(B)} = \tr_{B(A)}(\rho)$ denotes the reduced state on system A and B, respectively. Since the binarised distribution has different numbers of inputs and outputs compared to the multi-outcome distribution, the former and the latter belongs to different correlations spaces. Consequently, their Bell nonlocality must be detected via different Bell inequalities.

\subsection{Constructing Bell inequalitites for binarised distribution}
We now show how to construct Bell inequalitites specifically tailored for certifying Bell nonlocality of a binarised distribution generated by projective measurements. In general, correlations are Bell nonlocal if they cannot be written as  $p(a,b| x,y) = \sum_\lambda p(\lambda) p(a | x, \lambda) p(b | y, \lambda)$ for some probability distribution $p(\lambda)$ and response functions $p(a | x, \lambda)$ and $p(b | y, \lambda)$. Importantly, since the randomness of the LHV model can be absorb into $p(\lambda)$, we can w.l.g take the response functions to be deterministic. We indicate this by using the notation $D(a | x, \lambda)$ and $D(b | y,\lambda)$ for deterministic response functions. Moreover, because the set of local correlations forms a polytope, deciding whether a target distribution is nonlocal can be cast as a linear program as follows \cite{TavakoliSDP2024}
\begin{equation} \label{eq:dual_LHV}
\begin{aligned}
        \min_{\{c_{abxy}\}} \quad &1+ \sum_{abxy} c_{abxy} p(a,b\lvert x,y) \\
         \text{s.t} &\quad 1+ \sum_{abxy} c_{abxy} p(a,b\lvert x,y) = \frac{1}{| n_A n_B|} \sum_{abxy} c_{abxy} \\
          & \sum_{abxy} c_{abxy} D(a \lvert x, \lambda) D(b \lvert y, \lambda) \geq 0 \quad \forall \ \lambda
\end{aligned}
\end{equation}
for some coefficients $c_{abxy}$. Here, $|n_A|$ and $|n_B|$ denote the cardinality of the set of possible outcomes of Alice and Bob, respectively. The distribution $ p(a,b|x,y)$ is local if the objective value is no less than one. When this is the case, one can extract from the program the relevant Bell inequality as
\begin{equation}\label{eq:Bell_ineq}
    \mathcal{W} = \sum_{abxy} c_{abxy}p(a,b\lvert x,y)  \geq 0,
\end{equation}
which is satisfied by all local models but violated by the considered distribution. 

Now, using the linear program defined in Eq.~\eqref{eq:dual_LHV} we can easily generate a Bell inequality for any binarised distribution. In the case that Alice and Bob perform binary measurements with $d$ possible outcomes, the generic form of the Bell inequality associated with the binarised implementation takes the form
\begin{equation}\label{eq:dual_LHV_bell}
    \mathcal{W}_{\text{bin}} = \sum_{x,y = 1}^{2d} \sum_{a,b = 1}^2 c_{abxy} p(ab \lvert xy) \geq 0.
\end{equation}
Specifically, we denote the Bell inequalities tailored for the binarised CGLMP distribution, and for its related variant tailored for maximally entangled states, by
$\mathcal{I}_{\text{bin}}$ and $\mathcal{S}_{\text{bin}}$, respectively. 

\subsection{Binarsed inequalitites}
We analyse the binarised four-outcome CGLMP inequality and its related inequality tailored for maximally entangled states. To start with, the binarised CGLMP distribution emulated from the optimal multi-outcome CGLMP distribution yields 
\begin{equation}
\mathcal{I}_{\text{bin}} = -0.186. 
\end{equation}
However, by performing a convex search algorithm (see Appendix \ref{App:see_saw}) that optimise $\mathcal{I}_{\text{bin}}$ over quantum states and measurements of fixed dimension, larger violations of $\mathcal{I}_{\text{bin}}$ can be obtained already with a distribution based on two-qubit entanglement, see table \ref{table:w_bin_satwap}. The reason why this is possible stems from the fact that the nonlocality of the binarised CGLMP distribution decrease with the dimension \cite{Tavakoli_2025}. Thus, $\mathcal{I}_{\text{bin}}$ cannot be used to certify four-dimensional entanglement. This emphasis the importance of performing multi-outcome measurements. 

Lastly, we also study the distribution for the Bell inequality tailored for maximally entangled states. To this end, we find that this distribution yields 
\begin{equation}
\mathcal{S}_{\text{bin}} = -0.200.
\end{equation}
Using, the convex search method we again find that higher violations of the witness can be obtained for lower-dimensional systems. The result is summaries in table \ref{table:w_bin_satwap}.

\begin{table}[ht!]
\begin{tabular}{l|lll}
\hline
Bin. Ineq.     & 2     & 3     & 4     \\ \hline
$\mathcal{I}_{\text{bin}}$ & -0.2129 & -0.2575 & -0.2575 \\
$\mathcal{S}_{\text{bin}}$ & -0.2094& -0.2532 & -0.2532
\end{tabular}
\caption{We establish lower bounds on the optimal violation of Bell inequalities $\mathcal{I}_{\text{bin}}$ and $\mathcal{S}_{\text{bin}}$, for quantum states and measurements of dimension $d = 2,3,4$.}
\label{table:w_bin_satwap}
\end{table}

\section{See-saw algorithm}\label{App:see_saw}
We now discuss a method, known as the see-saw algorithm, which can be used to establish lower bounds on the optimal quantum violation of an arbitrary Bell inequality
\begin{equation}
\mathcal{W} = \sum_{abxy} c_{abxy} p(a,b\mid x,y).
\end{equation}
From Born's rule, we have that the quantum correlations are given by $ p(a,b| x,y) = \tr(A_{a \mid x} \otimes B_{b\mid y} \rho)$. Here, $\{A_{a\mid x}\}_a$ and $\{B_{b \mid y}\}_b$ corresponds to Alice's and Bob's measurements, respectively, and $\rho$ denote the bipartite state shared between the parties. The see-saw method is an alternating convex search algorithm and the main idea of this method is to systematically increase the accuracy of a lower bound to $\mathcal{W}$, by optimising the function over multiple convex programs iteratively, until convergence is reached. 

Specifically, the see-saw algorithm goes as follows: Start by sampling a random state $\rho$ and random measurements $\{B_{b\mid y}\}_b$. Define the SDP variable to be $\{A_{a\mid x}\}_a$ and optimise the Bell inequality over these operators
\begin{equation}\label{eq:see_saw_1}
\begin{aligned}
\max_{\{A_{a \mid x}\}_a} \quad &\mathcal{W} = \sum_{a,b,x,y} c_{abxy} \tr(A_{a \mid x} \otimes B_{b\mid y} \rho)\\
\text{s.t}. \quad & \sum_a A_{a\mid x} = \openone \quad \forall \ x \\
& A_{a \mid x} \succeq 0 \quad \forall \ a,x.
\end{aligned}
\end{equation}
Thereafter, fix Alice's measurements to be the SDP variables $\{ A_{a\mid x}^*\}_a$ that maximises the objective function. Next, $\mathcal{W}$ is optimised with respect to Bob's measurements, with a program analogous to the one in Eq. \eqref{eq:see_saw_1}. Thereafter, fixing Bob's measurement to be the optimal SDP variables $\{B_{b\mid y}^*\}_b$, we maximise the objective function with respect to the quantum state $\rho$ as follows
\begin{equation}\label{eq:see_saw_2}
\begin{aligned}
\max_{\rho} \quad &\mathcal{W}= \sum_{a,b,x,y} c_{abxy} \tr(A_{a \mid x} \otimes B_{b\mid y} \rho)\\
\text{s.t}. \quad & \tr(\rho) = 1, \quad \text{and} \quad \rho \succeq 0.
\end{aligned}
\end{equation}
The three optimisation programs are then iterated until $\mathcal{W}$ converges to a local maximum, yielding a lower bound on the optimal quantum value of $\mathcal{W}$.

\section{Details of multi-outcome measurement}\label{sec:exp_measurement}

\begin{figure}[htb!]
	\includegraphics[width=0.75\linewidth]{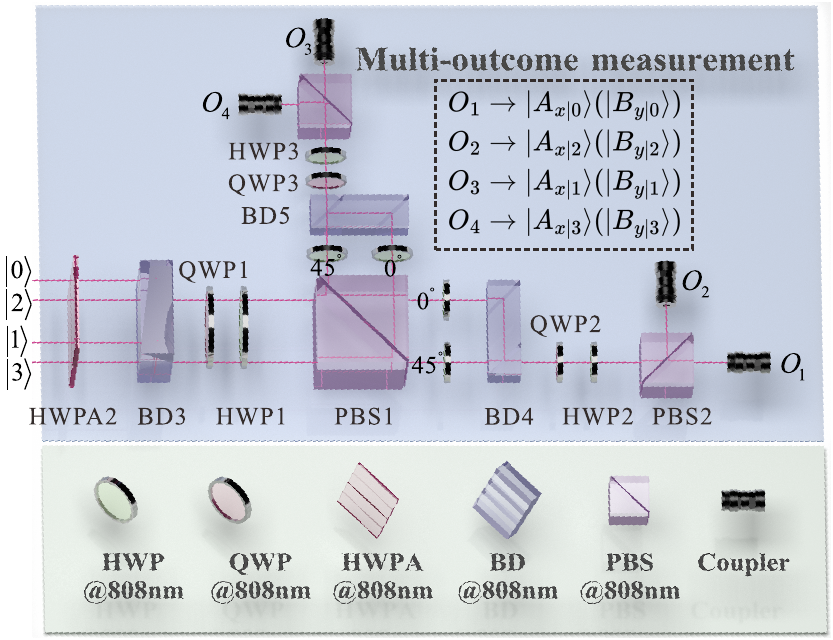}
	\vspace{-0.3cm}%
	\caption{Measurement setup for Alice and Bob. Input state is encoded in $2\times2$ path d.o.f., $\ket{0}$, $\ket{1}$ for upper two paths and $\ket{2}$ $\ket{3}$ for lower two, converted into path-polarization hybrid state with polarization adjustment and BD3-5. The BD3 refracts horizontally polarized beams downwards by $4~mm$, while BD4 and BD5 refract vertically polarized beams horizontally by $4~mm$. Projection is achieved by HWP1-3 and QWP1-3 with following PBS, resulting in photon counts for $O_1 - O_4$, each of which corresponds to a different 4-dimensional projective measurement as listed in the box.}
	\label{fig:measurement_sup}
	\vspace{0.2cm}
\end{figure}

Our multi-outcome measurement setup for four-dimensional path-entangled photons is designed to realize arbitrary projective measurements in a four-dimensional Hilbert space without resorting to sequential binary-outcome detections. The setup, shown in Fig. \ref{fig:setup} (b) and (c) of the main text, also detailed in Fig. \ref{fig:measurement_sup}, performs a single-shot four-outcome measurement by transforming path-encoded states into a hybrid path-polarisation encoding and then projecting onto four spatially separated output modes.

The measurement proceeds in two sequential projection steps, each acting on a two-dimensional subspace. A combination of a quarter-wave plate (QWP), a half-wave plate (HWP), and a polarising beam splitter (PBS3) is used to project onto superpositions of the two polarisation-distinguished subspaces. By setting the angles of the QWP and HWP appropriately, we select the relative phases $\varphi_{02}$ and $\varphi_{13}$ between the states $\{\ket{0},\ket{2}\}$ and $\{\ket{1},\ket{3}\}$. Photons transmitted through PBS3 correspond to the subspace $\{ \ket{0} + e^{i\varphi_{02}} \ket{2}, \ket{1} + e^{i\varphi_{13}} \ket{3} \}$, while reflected photons correspond to $\{ \ket{0} - e^{i\varphi_{02}} \ket{2}, \ket{1} - e^{i\varphi_{13}} \ket{3} \}$. These two output paths remain separated by 4 mm horizontally.

\begin{table}[b!]
\centering
\caption{Measurement setting of Alice and Bob for $\mathcal{I}_4$.}
\begin{tabular}{ccccccc}
\hline\hline
Measurement & QWP1 & HWP1 & QWP2 & HWP2 & QWP3 & HWP3 \\ 
\hline
$A_1$ & $45^\circ$ & $22.5^\circ$ & $45^\circ$ & $22.5^\circ$ & $45^\circ$ & $0^\circ$ \\
$A_2$ & $45^\circ$ & $0^\circ$ & $45^\circ$ & $33.75^\circ$ & $45^\circ$ & $11.25^\circ$ \\
$B_1$ & $45^\circ$ & $33.75^\circ$ & $45^\circ$ & $16.875^\circ$ & $45^\circ$ & $39.375^\circ$ \\
$B_2$ & $45^\circ$ & $11.25^\circ$ & $45^\circ$ & $28.125^\circ$ & $45^\circ$ & $50.625^\circ$ \\
\hline\hline
\label{table:exp_cglmp_setting}
\end{tabular}
\end{table}

\begin{table}[b!]
\centering
\caption{Measurement setting of Alice and Bob for $\mathcal{S}_4$.}
\begin{tabular}{ccccccc}
\hline\hline
Measurement & QWP1 & HWP1 & QWP2 & HWP2 & QWP3 & HWP3 \\ 
\hline
$A_1$ & $45^\circ$ & $33.75^\circ$ & $45^\circ$ & $16.875^\circ$ & $45^\circ$ & $84.375^\circ$ \\
$A_2$ & $45^\circ$ & $56.25^\circ$ & $45^\circ$ & $5.625^\circ$ & $45^\circ$ & $73.125^\circ$ \\
$B_1$ & $45^\circ$ & $0^\circ$ & $45^\circ$ & $33.75^\circ$ & $45^\circ$ & $56.25^\circ$ \\
$B_2$ & $45^\circ$ & $-22.5^\circ$ & $45^\circ$ & $45^\circ$ & $45^\circ$ & $-22.5^\circ$ \\
\hline\hline
\label{table:exp_satwap_setting}
\end{tabular}
\end{table}

Each of the two output paths from the first stage is further processed by another set of BDs (BD4 and BD5), QWP, HWP, and PBS. This stage measures the relative phases between $\{\ket{0},\ket{1}\}$ and $\{\ket{2},\ket{3}\}$. After this step, the photons are directed into one of four distinct output paths, each corresponding to one of the four basis states of the target measurement operator.

The four output paths are coupled into single-mode fibers ($O_1 - O_4$) connected to single-photon detectors. All four outcomes are registered simultaneously in each experimental round. No post-selection is applied: the measured counts for each outcome are normalized directly among the four detectors, ensuring that the normalization condition $\sum_a \ket{A_{a|x}} = \openone$
 is not assumed but physically enforced. This closes the binarisation loophole.

Alice and Bob can actively choose the measurement bases by adjusting the angle of wave plates and decide which base corresponds to a specific coupler. The angles for the measurement of $I_4$ and $S_4$ are listed in Table. \ref{table:exp_cglmp_setting} and Table \ref{table:exp_satwap_setting}.

\section{Detailed experimental data}\label{sec:exp_data}

Our data of $\mathcal{I}_4$ and $\mathcal{S}_4$ is listed below in Table. \ref{tab:exp_results_cg} and Table. \ref{tab:exp_results_sat} respectively. Each $4\times4$ block is collected simultaneously by Alice and Bob with their multi-outcome measurement of a specific setting, and then normalised. We have coincidence counting of ~400 Hz in total, ~100 Hz for each subspace, the data are collected in 100s for each setting. Thus we have 
\begin{equation}
	\begin{aligned}
		& \mathcal{I}_4^{\text{exp}}=0.3346\pm0.0030,\\ &\mathcal{S}_4^{\text{exp}}=0.2832\pm0.0027,
	\end{aligned}
	\label{eq:expresult_sup}
\end{equation}
where the standard deviation is estimated from Poisson statistics of photon counting through 10000 Monte-Carlo simulations. Also the data is listed in Fig.\ref{fig:resultsbar} for a more intuitive view.

\begin{table*}[t]
    \centering
    \renewcommand\arraystretch{1.3}
    \caption{Experimental joint probability distributions for $\mathcal{I}_4$. Each $4\times 4$ block is obtained from normalisation of the coincidence counts in one round of multi-outcome measurement.}
    \label{tab:exp_results_cg}
    \setlength{\tabcolsep}{6pt}
    \begin{tabular}{c|c c c c| c c c c c}
    \hline
    
    $\mathcal{I}_4$ & A1=0 & A1=1 & A1=2 & A1=3   & A2=0 & A2=1 & A2=2 & A2=3 \\
    \hline
    B1=0 & 0.1854 & 0.0295 & 0.0093 & 0.0041 &  0.2030 & 0.0081 & 0.0011 & 0.0236 \\
    B1=1 & 0.0055 & 0.2146 & 0.0282 & 0.0322 &  0.0193 & 0.2408 & 0.0060 & 0.0075 \\
    B1=2 & 0.0057 & 0.0045 & 0.1802 & 0.0244 &  0.0016 & 0.0240 & 0.1760 & 0.0073 \\
    B1=3 & 0.0257 & 0.0301 & 0.0075 & 0.2132 &  0.0106 & 0.0048 & 0.0263 & 0.2401 \\
    \hline
    B2=0 & 0.1967 & 0.0228 & 0.0008 & 0.0116 &  0.0347 & 0.0119 & 0.0085 & 0.2137 \\
    B2=1 & 0.0090 & 0.2283 & 0.0249 & 0.0168 &  0.2010 & 0.0329 & 0.0093 & 0.0133 \\
    B2=2 & 0.0028 & 0.0032 & 0.1749 & 0.0322 &  0.0050 & 0.1888 & 0.0255 & 0.0068 \\
    B2=3 & 0.0234 & 0.0157 & 0.0080 & 0.2288 &  0.0215 & 0.0127 & 0.1821 & 0.0323 \\
    \hline
\end{tabular}
\end{table*}

\begin{table*}[t]
    \centering
    \renewcommand\arraystretch{1.3}
    \caption{Experimental joint probability distributions for $\mathcal{S}_4$. Each $4\times 4$ block is obtained from normalisation of the coincidence counts in one round of multi-outcome measurement.}
    \label{tab:exp_results_sat}
    \setlength{\tabcolsep}{6pt}
    \begin{tabular}{c|c c c c| c c c c c}
    \hline
    
    $\mathcal{S}_4$ & A1=0 & A1=1 & A1=2 & A1=3   & A2=0 & A2=1 & A2=2 & A2=3 \\
    \hline
    B1=0 & 0.1814 & 0.0163 & 0.0190 & 0.0267 &  0.1968 & 0.0273 & 0.0013 & 0.0082 \\
    B1=1 & 0.0321 & 0.1952 & 0.0036 & 0.0079 &  0.0184 & 0.1962 & 0.0189 & 0.0105 \\
    B1=2 & 0.0209 & 0.0306 & 0.2097 & 0.0144 &  0.0033 & 0.0086 & 0.2329 & 0.0314 \\
    B1=3 & 0.0047 & 0.0055 & 0.0344 & 0.1976 &  0.0151 & 0.0072 & 0.0225 & 0.2015 \\
    \hline
    B2=0 & 0.0060 & 0.0111 & 0.0325 & 0.2180 &  0.1890 & 0.0167 & 0.0165 & 0.0239 \\
    B2=1 & 0.1757 & 0.0082 & 0.0083 & 0.0258 &  0.0281 & 0.1913 & 0.0053 & 0.0046 \\
    B2=2 & 0.0313 & 0.2220 & 0.0068 & 0.0098 &  0.0165 & 0.0266 & 0.2157 & 0.0168 \\
    B2=3 & 0.0114 & 0.0244 & 0.2007 & 0.0080 &  0.0064 & 0.0041 & 0.0309 & 0.2078 \\
    \hline
\end{tabular}
\end{table*}

\begin{figure*}[ht]
	\includegraphics[width=\textwidth]{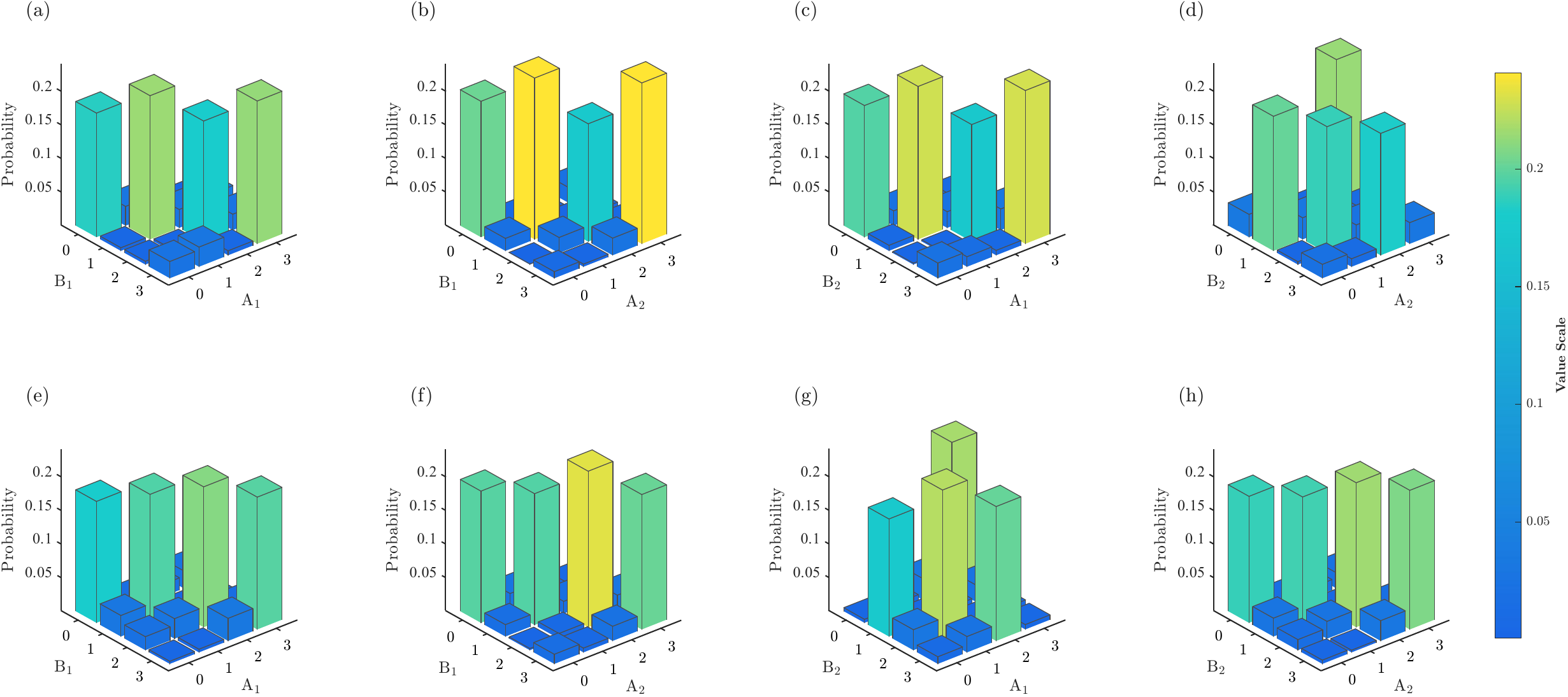}
	\vspace{-0.3cm}%
	\caption{Experimental results. (a)-(d) are measurement results of $\mathcal{I}_4$, and (e)-(h) are results of $\mathcal{S}_4$. Each of the plot is normalized with respect to the data of four different outcomes. The axis labeled by $A_x$ and $B_y$ stands for Alice's and Bob's measurement result, from 0 to 3. Each bar can be mapped to the joint probability of their measurement result, respect to the color bar.}
	\label{fig:results}
	\vspace{0.2cm}
\end{figure*}

\section{Finite count statistics}\label{App:Counts}
The experimental violation of the four-outcome CGLMP is based on estimating probabilities from finite statistics. Therefore, to be confident that an observed violation of these inequalities is due genuine-four-dimensional entanglement, and is not caused by statistical fluctuations, we perform a finite count statistics analysis. The analysis is done under the null-hypothesis, which test how likely it is that an experimentally observed effect is due to chance alone. 

Consider the four-outcome CGLMP inequality $\mathcal{I}_4$. To use the null-hypothesis we start by normalising the Bell test such that the maximal violation attained by ququart entanglement is given by $\hat{\mathcal{I}}_4 = 1$. We now define the normalised violation gap between the experimentally measured violation $\hat{\mathcal{I}}_4^{\text{exp}} = 0.9174$ and the optimal violation achievable with qutrit-entanglement $\hat{\mathcal{I}}_4^{(3)} = 0.8356$ as follows $\hat{\Delta}_{\text{exp}} = \mathcal{I}_4^{\text{exp}}- \hat{\mathcal{I}}_4^{(3)} = 0.0813$. We now make use of Chernoff's bound, which states that the probability that a random violation $\Delta$ exceeds the experimentally measured violation is upper bounded by \cite{Chernoff}
\begin{equation}
\text{P}( \Delta \geq \Delta_{\text{exp}}) \leq e^{-D_{KL}(\hat{\mathcal{I}}_4^{(3)}+\Delta_{\text{exp}} \parallel \hat{\mathcal{I}}_4^{(3)})N }
\end{equation}
where $N$ is the total number of measurements and $D_{KL}(x \parallel y)$ is the Kullback-Liebler divergence
\begin{equation}
D_{KL}(x \parallel y) = x \ln\left(\dfrac{x}{y}\right)+(1-x)\ln \left( \dfrac{1-x}{1-y}\right).
\end{equation}
The probability $\text{P}( \Delta \geq \Delta_{\text{exp}})$ can then be interpreted as the $p$-value of our measurement, i.e., the probability of obtaining a result at least as extreme, given that the null-hypothesis is true. Given that we have chosen $p$-values that we find acceptable for the observed violation $\Delta_{\text{exp}}$, we can compute a lower bound on the counts $N$ required to achieve this condition
\begin{equation}
N > \dfrac{1}{D_{KL}( \hat{\mathcal{I}}_4^{(3)}+\Delta_{\text{exp}} \parallel \hat{\mathcal{I}}_4^{(3)})}\operatorname{ln}\bigg(\dfrac{1}{\operatorname{P}(\Delta \geq \Delta_{\text{exp}})}\bigg).
\end{equation}

We now want to estimate with what probability that we can reject the null-hypothesis that the experimental violation of the CGLMP inequality is due to statistical fluctuations. In this case, roughly $N \simeq 2420$ coincidences results in a $p$-value of order $10^{-30}$. With $N \simeq 24,000$ coincidences $p$-value of order $10^{-300}$. Therefore, since the total number of counts in the experiment is $N \simeq 200,000$, we can with very high certainty reject the null-hypothesis since the $p$-value is negligible.

Next, we perform a similar analysis for the related CGLMP inequality, tailored for maximally entangled state. Renormalising the inequality such that $ \hat{\mathcal{S}}_4 = 1$, the optimal bound achievable with qutrit entanglement reads $\hat{\mathcal{S}}_4^{(3)} = 0.9571$, whereas the experimental violation is $\hat{\Delta}_{\text{exp}} = 0.0342$. In this case, roughly $N \simeq 33000$ coincidences results in a $p$-value of order $10^{-300}$. Since the total number of counts in the experiment is $N \simeq 170,000$, we can with high certainty reject the null-hypothesis.

\end{document}